\documentclass[aps,preprint,showpacs,superscriptaddress]{revtex4}
\newcommand{\beq}{\begin{equation}}
\newcommand{\eeq}{\end{equation}}
\begin{document}

\title{Quasiparticles and order parameter near quantum phase transition
in heavy fermion metals}

\author{V.R.  Shaginyan \footnote{E--mail:
vrshag@thd.pnpi.spb.ru}} \affiliation{Petersburg Nuclear Physics
Institute, Russian Academy of Sciences,  Gatchina, 188300, Russia}
\affiliation{ CTSPS, Clark Atlanta University, Atlanta, Georgia
30314, USA}
\author{A.Z. Msezane}\affiliation{ CTSPS, Clark Atlanta University,
Atlanta, Georgia 30314, USA}
\author{M.Ya. Amusia}\affiliation{ Racah Institute of Physics,
the Hebrew University, Jerusalem 91904, Israel} \affiliation{ A.F.
Ioffe Physical-Technical Institute, Russian Academy of Sciences,
St. Petersburg, 194021, Russia}

\begin{abstract}

It is shown that the Landau paradigm based upon both the
quasiparticle concept and the notion of the order parameter is
valid and can be used to explain the anomalous behavior of the
heavy fermion metals near quantum critical points. The
understanding of this phenomenon has been problematic largely
because of the absence of theoretical guidance. Exploiting this
paradigm and the fermion condensation quantum phase transition, we
investigate the anomalous behavior of the heavy electron liquid
near its critical point at different temperatures and applied
magnetic fields. We show that this anomalous behavior is universal
and can be used to capture the essential aspects of recent
experiments on heavy-fermion metals at low temperatures.

\end{abstract}


\maketitle

Recent experiments have demonstrated that at low temperatures the
main properties of the heavy-fermion (HF) metals such as the
magnetoresistance, resistivity, specific heat, magnetization,
susceptibility, volume thermal expansion, etc, strongly depend on
temperature $T$  and applied magnetic field $B$. As a result,
these properties can be controlled by placing these metals at the
special point of the field-temperature $B-T$ phase diagram. In the
Landau Fermi liquid (LFL) theory, considered as the main
instrument when investigating quantum many electron physics, the
effective mass $M^*$ of quasiparticle excitations determining the
thermodynamic properties of electronic systems is practically
independent of the temperature and applied magnetic fields.
Therefore, the observed anomalous behavior is uncommon and can be
hardly understood within the framework of the LFL theory.
Consequently, it is theories necessary to invoke that are based on
the Landau concept of the order parameter which is introduced  to
classify phases of matter. These theories associate the anomalous
behavior with critical fluctuations of the magnetic order
parameter. These fluctuations suppressing the quasiparticles are
attributed to a conventional quantum phase transition taking place
when the system in question approaches its quantum critical point
(QCP). It has become generally accepted that the fundamental
physics that gives rise to the high-$T_c$ superconductivity and
non-Fermi liquid (NFL) behavior, with the recovery of the
Landau-Fermi liquid (LFL) behavior under the application of
magnetic fields observed in the HF metals and high-$T_c$
compounds, is controlled by quantum phase transitions. This has
made quantum phase transitions a subject of intense current
interest, see e.g. \cite{sac,voj}.

A quantum phase transition is driven by control parameters such as
composition, pressure, number density $x$ of electrons (holes),
magnetic field $B$, etc, and takes place at QCP when the
temperature $T=0$. QCP separates an ordered phase generated by
quantum phase transition from a disordered phase. It is expected
that the universal behavior is only observable if the heavy
electron liquid in question is very near QCP, for example, when
the correlation length is much larger than the microscopic length
scales. Quantum phase transitions of this type are quite common,
and we shall label them as conventional quantum phase transitions
(CQPT). In the case of CQPT, the physics is dominated by thermal
and quantum fluctuations of the critical state, which is
characterized by the absence of quasiparticles. It is believed
that the absence of quasiparticle-like excitations is the main
cause of the NFL behavior and other types of the critical behavior
in the quantum critical region. However, theories based on CQPT
fail to explain the experimental observations related to the
divergence of the effective mass $M^*$ at the magnetic field tuned
QCP, the specific behavior of the spin susceptibility and its
scaling properties, the thermal expansion behavior, etc, see e.g.
\cite{geg,geg1,pag1,takah,bi,cust,bud,cap,
pag,kuch,movsh,senth1,senth2,senth3,zhu}.

The LFL theory rests on the notion of quasiparticles which
represent elementary excitations of a Fermi liquid. Therefore
these are appropriate excitations to describe the low temperature
thermodynamic properties. In the case of an electron system, these
are characterized by the electron's quantum numbers and effective
mass $M^*$ independent of the temperature and magnetic field. The
inability of the LFL theory to explain the experimental
observations which point to the dependence of $M^*$ on the
temperature $T$ and applied magnetic field $B$ may lead to the
conclusion that the quasiparticles do not survive near QCP, and
one might be further led to the conclusion that the heavy electron
does not retain its integrity as a quasiparticle excitation, see
e.g. \cite{cust,senth1,senth2,senth3}.

The mentioned above inability to explain the behavior of the HF
metals at QCP within the framework of theories  based on CQPT may
also lead to the conclusion that the other important Landau
concept of the order parameter fails as well, see e.g.
\cite{senth1,senth2,senth3}. Thus, we are left without the most
fundamental principles of many body quantum physics while a great
deal of interesting NFL phenomena related to the anomalous
behavior and the experimental facts collected in measurements on
the HF metals remain out of reasonable theoretical explanations.

In this Letter we show that the Landau paradigm resting upon both
the quasiparticle concept and the notion of the order parameter
are valid and can be used to explain the anomalous behavior of the
HF metals near QCP. Exploiting this paradigm, we identify QCP
observed in the HF metals as the QCP corresponding to the fermion
condensation quantum phase transition (FCQPT) and consider the
behavior of the quasiparticle effective mass $M^*(T,B)$ as a
function of temperature $T$, applied magnetic field $B$ and the
number density $x$. This behavior is universal and can be used to
explain the main properties of the HF metals at low temperatures
such as the magnetoresistance, resistivity, specific heat,
magnetization, volume thermal expansion, etc. We also demonstrate
that the theory captures the essential aspects of recent
experiments in the HF metals at low temperatures.

Assuming the existence of quasiparticles, we start with the Landau
equation determining $M^*(T,B)$ in the case of a homogeneous
liquid at finite temperatures and low magnetic field, see e.g.
\cite{lanl1} \beq \frac{1}{M^*(T,B)}=
\frac{1}{M}+\sum_{\sigma_1}\int \frac{{\bf p}_F{\bf p_1}}{p_F^3}
F_{\sigma,\sigma_1}({\bf p_F},{\bf p}_1) \frac{\partial
n_{\sigma_1}({\bf p}_1,T,B)}{\partial {p}_1} \frac{d{\bf
p}_1}{(2\pi)^3}. \eeq Here $F_{\sigma,\sigma_1}({\bf p_F},{\bf
p}_1)$ is the Landau amplitude depending on the momenta $p$ and
spins $\sigma$, $p_F$ is the Fermi momentum, $M$ is the bare mass
of an electron and $n_{\sigma}({\bf p},T)$ is the quasiparticle
distribution function. Since HF metals are predominantly three
dimensional (3D) structures we treat the homogeneous heavy
electron liquid as a $3$D liquid also. For the sake of simplicity,
we omit the spin dependence of the effective mass since in the
case of a homogeneous liquid and weak magnetic fields, $M^*(T,B)$
does not noticeably depend on the spins. The quasiparticle
distribution function is of the form
\begin{equation} n_{\sigma}({\bf p},T)=\left\{ 1+\exp
\left[ \frac{(\varepsilon({\bf p},T)-\mu_{\sigma})}T \right]
\right\} ^{-1},
\end{equation}
where $\varepsilon({\bf p},T)$ is the single-particle spectrum, or
dispersion, of the quasiparticle excitations and $\mu_{\sigma}$ is
the chemical potential. The single-particle spectrum is given by
the equation \cite{lanl1}
\begin{equation}
\varepsilon({\bf p},T)=\frac{\delta E[n(p)]}{\delta n({\bf p},T)},
\end{equation}
where $E[n(p)]$ is the Landau functional defining the ground state
energy $E$. In our case, the single-particle spectrum does not
noticeably depend on the spin, while the chemical potential may
have a dependence due to the Zeeman splitting. We will show
explicitly the spin dependence of a physical value when this
dependence is of importance for understanding. Since
$\varepsilon(p\sim p_F,T)-\mu\sim p_F(p-p_F)/M^*$, it follows from
Eq. (2) that the quasiparticle distribution function becomes the
step function of the momentum, $n(p,T=0)=\theta(p_F-p)$, at zero
temperature, provided that the effective mass is positive and
finite. Applying Eq. (1) at $T=0$ and $B=0$ and taking into
account that $n(p,T=0)$ becomes the step function we obtain the
standard result
$$ \frac{M^*}{M}=\frac{1}{1-N_0F^1(p_F,p_F)/3}.$$
Here $N_0$ is the density of states of the free Fermi gas and
$F^1(p_F,p_F)$ is the $p$-wave component of the Landau
interaction. Since in the LFL theory $x=p_F^3/3\pi^2$, the Landau
amplitude can be written as $F^1(p_F,p_F)=F^1(x)$. Assume that at
some critical point $x_{FC}$ the denominator
$(1-N_0F^1(p_F,p_F)/3)$ tends to zero, that is
$(1-N_0F^1(x)/3)\propto(x-x_{FC})\to 0$. As a result, one obtains
that $M^*(x)$ behaves as \cite{shag1,khod1}
\begin{equation}
\frac{M^*(x)}{M}\propto\frac{x_{FC}}{x-x_{FC}}\propto\frac{1}{r}.
\end{equation}
Here $r=(x-x_{FC})$ is the ``distance'' from the QCP of FCQPT
taking place at $x_{FC}$. The observed behavior is in good
agreement with recent experimental observations, see e.g.
\cite{skdk,cas1}, and calculations \cite{krot,sarm1,sarm2}. In the
case of electronic systems Eq. (4) is valid at low densities
\cite{ksk,ksz}. Such a behavior of the effective mass can be
observed in the HF metals with a quite flat, narrow conduction
band, corresponding to a large effective mass $M^*(x\simeq
x_{FC})$, with strong electron correlations and the effective
Fermi temperature $T_k\sim p_F^2/M^*(x)$ of the order of a few
Kelvin or even lower  \cite{stew}.

Replacing $n_{\sigma}({\bf p},T,B)$ by $n_{\sigma}({\bf
p},T,B)\equiv \delta n_{\sigma}({\bf p},T,B)+n_{\sigma}({\bf
p},T=0,B=0)$ where $\delta n_{\sigma}({\bf p},T,B)=
n_{\sigma}({\bf p},T,B)-n_{\sigma}({\bf p},T=0,B=0)$, Eq. (1)
takes the form \beq \frac{M}{M^*(T,B)}=
\frac{M}{M^*(x)}+\frac{M}{p_F^2}\sum_{\sigma_1}\int \frac{{\bf
p}_F{\bf p_1}}{p_F} F_{\sigma,\sigma_1}({\bf p_F},{\bf p}_1)
\frac{\partial \delta n_{\sigma_1}({\bf p}_1,T,B)}{\partial {p}_1}
\frac{d{\bf p}_1}{(2\pi)^3}. \eeq Note, that $M^*(x)$ is given by
Eq. (4). In the case of normal metals with the effective mass of
the order of a few bare electron masses and up to temperatures $T
\sim 10$ K, the second term on the right hand side of Eq. (5) is
of the order of $T^2/\mu^2$ and is much smaller than the first
term. Thus, the system in question demonstrates the LFL behavior
with the effective mass being practically independent of
temperature, that is the corrections are proportional to $T^2$.
One can check that the same is true when magnetic field up to
$B\sim 10$ T is applied. Near the critical point $x_{FC}$, when
$M/M^*(x\to x_{FC})\to 0$, the behavior of the effective mass
changes drastically because the first term vanishes and the second
term determines the effective mass itself rather than small
corrections to $M^*(x)$ related to $T$ and $B$. In that case, Eq.
(5) no longer explicitly depends on $M^*(x)$ and determine the
effective mass as a function of $B$ and $T$. As we will see, Eq.
(5) describes both the NFL behavior and the LFL one with the
presence of quasiparticles. In contrast to the conventional
quasiparticles these are characterized by an effective mass that
strongly depends on both the magnetic field and the temperature.

Let us turn to a qualitative analysis of the solutions of Eq. (5)
when $x\simeq x_{FC}$. We start with the case when $T=0$ and $B$
is finite. The application of magnetic field leads to the Zeeman
splitting of the Fermi surface and the difference $\delta p$
between the Fermi surfaces with ``spin up'' and ``spin down''
becomes  $\delta p=p_F^{\uparrow}-p_F^{\downarrow}\sim
\mu_{B}BM^*(B)/p_F$ with $\mu_{B}$ being the Bohr magneton. Upon
taking this into account, we observe that the second term in Eq.
(5) is proportional to $(\delta p)^2 \propto
(\mu_{B}BM^*(B)/p_F)^2$, and Eq. (5) takes the form \beq
\frac{M}{M^*(B)}\simeq
\frac{M}{M^*(x)}+c\frac{(\mu_{B}BM^*(B))^2}{p_F^4}, \eeq where $c$
is a constant. Note that the effective mass $M^*(B)$ depends on
$x$ as well and this dependence disappears at $x=x_{FC}$. At the
point $x=x_{FC}$, the term $M/M^*(x)$ vanishes, Eq. (6) becomes
uniform and can be solved analytically \cite{shag1,ckhz} \beq
\frac{M}{M^*(B)}\propto \frac{1}{(B-B_{c0})^{2/3}}. \eeq Here,
$B_{c0}$ is the critical magnetic field which drives both a HF
metal to its magnetic field tuned QCP and the corresponding N\'eel
temperature toward $T=0$ \cite{shag4}. We note that in some cases
$B_{c0}=0$, for example, the HF metal CeRu$_2$Si$_2$ shows neither
evidence of the magnetic ordering, superconductivity down to the
lowest temperatures nor the LFL behavior  \cite{takah}. 
When deriving Eq. (6), we have omitted the next terms on the right hand side  of Eq. (6) assuming that  $(B\mu_BM/p_F^2)\ll 1$. At
$x>x_{FC}$, $M^*(x)$ is finite and we are dealing with the
conventional Landau quasiparticles provided that the magnetic
field is weak, so that $M^*(x)/M^*(B)\ll 1$ with $M^*(B)$ is given
by Eq. (7). In that case, the second term on the right hand side
of Eq. (6) is proportional to $(BM^*(x))^2$ and represents small
corrections. In the opposite case, when $M^*(x)/M^*(B)\gg 1$, the
heavy electron liquid behaves as it were placed at QCP. Since in
the LFL regime the main thermodynamic properties of the system is
determined by the effective mass, it follows from Eq. (7) that we
obtain a unique possibility to control the magnetoresistance,
resistivity, specific heat, magnetization, volume thermal
expansion, etc. At this point, we note that the large effective
mass leads to the high density of states provoking a large number
of states and phase transitions to emerge and compete with one
another. Here we assume that these can be suppressed by the
application of a magnetic field and concentrate on the
thermodynamical properties.

To consider the qualitative behavior of $M^*(T)$ at elevated
temperatures, we simplify Eq. (5) by omitting the variable $B$ and
mimicking the influence of the applied magnetic field by the
finite effective mass entering the denominator of the first term
on the right hand side of Eq. (5). This effective mass becomes a
function of the distance $r$, $M^*(r)$, which  is determined by
both $B$ and $(x-x_{FC})$. If the magnetic field vanishes the
distance is $r=(x-x_{FC})$. We integrate the second term over the
angle variable, then over $p_1$ by parts and substitute the
variable $p_1$ by $z$, $z=(\varepsilon(p_1)-\mu)/T$. In the case
of the flat and narrow band, we use an approximation
$(\varepsilon(p_1)-\mu)\simeq p_F(p_1-p_F)/M^*(T)$ and obtain \beq
\frac{M}{M^*(T)}=\frac{M}{M^*(r)} +\alpha\int^{\infty}_{0}
F(p_F,p_F(1+\alpha z)) \frac{1}{1+e^z}dz
-\alpha\int^{1/\alpha}_{0} F(p_F,p_F(1-\alpha z))
\frac{1}{1+e^z}dz. \eeq Here the function $F\sim M d(F^1p^2)/dp$,
the factor $\alpha=TM^*(T)/p_F^2= TM^*(T)/(T_kM^*(r))$, and the
Fermi momentum is defined as $\varepsilon(p_F)=\mu$. We first
assume that $\alpha\ll 1$. Then omitting terms of the order of
$\exp(-1/\alpha)$, we expand the upper limit of the second
integral on the right hand side of Eq. (8) to $\infty$ and observe
that the sum of the second and third terms represents an even
function of $\alpha$. These are the typical integrals containing
the Fermi-Dirac function and can be calculated by using  standard
procedures, see e.g. \cite{lanl2}. Since we need only an
estimation of the integrals, we represent Eq. (8) as \beq
\frac{M}{M^*(T)}\simeq
\frac{M}{M^*(r)}+a_1\left(\frac{TM^*(T)}{T_kM^*(r)}\right)^2
+a_2\left(\frac{TM^*(T)}{T_kM^*(r)}\right)^4+...\eeq Here $a_1$
and $a_2$ are constants of the order of units. Equation (9) can be
regarded as a typical equation of the LFL theory with the only
exception being the effective mass $M^*(r)$ which strongly depends
on the distance $r=x-x_{FC}\geq 0$ and diverges as $r\to 0$.
Nonetheless, it follows from Eq. (9) that when $T\to 0$, the
corrections to $M^*(r)$ start with the $T^2$ terms provided that
\beq M/M^*(r)\gg \left(\frac{TM^*(T)}{T_kM^*(r)}\right)^2\simeq
\frac{T^2}{T_k^2}, \eeq and the system exhibits the LFL behavior.
It is seen from Eq. (10) that when $r\to 0$,  $M^*(r)\to\infty$,
and the LFL behavior expires. The free term on the right hand side
of Eq. (8) vanishes, $M/M^*(r)\to 0$, and Eq. (8) in itself
determines the value and universal behavior of the effective mass.

At some temperature $T_1\ll T_k$, the value of the sum on the
right hand side of Eq. (9) is determined by the second term. Then
Eq. (10) is not valid, and upon keeping only the second term in
Eq. (9) this can be used to determine $M^*(T)$ in a transition
region \cite{ckhz,shag5} \beq M^*(T)\propto \frac{1}{T^{2/3}}.\eeq
The evolution of the $-2/3$ exponent as the temperature increases
is worth a comment. Equation (11) is valid if the second term in
Eq. (9) is much larger than the first one, that is \beq
\frac{T^2}{T_k^2} \gg \frac{M}{M^*(r)},\eeq and this term is
grater than the third one, \beq \frac{T}{T_k} \ll
\frac{M^*(r)}{M^*(T)}\simeq 1.\eeq Obviously, both Eqs. (12) and
(13) can be simultaneously satisfied if $M/M^*(r)\ll 1$ and $T$ is
finite. The range of temperatures over which Eq. (11) is valid
shrinks to zero as soon as $r\to 0$ because $T_k\to 0$. Thus, if
the system is very near QCP, $x\to x_{FC}$, it is possible to
observe the behavior of the effective mass given by Eq. (11) in a
wide range of temperatures provided that the effective mass
$M^*(r)$ is diminished by the application of high magnetic fields,
that is, the distance $r$ becomes larger due to $B$. When $r$ is
finite the $T^{-2/3}$ behavior can be observed at relatively high
temperatures. To estimate the transition temperature $T_1$, we
observe that the effective mass is a continuous function of the
temperature, thus $M^*(B)\sim M^*(T_1)$. Taking into account Eqs.
(7) and (11), we obtain $T_1\propto B$.

Then, at elevated temperatures, the system enters into a different
regime. The coefficient $\alpha$ becomes $\alpha\sim 1$, the upper
limit of the second integral in Eq. (8) cannot be expanded to
$\infty$, and odd terms come into play. As a result, Eq. (9) is no
longer valid, but the sum of both the first integral and the
second one on the right hand side of Eq. (8) is proportional to
$M^*(T)T$. Upon omitting the first term $M/M^*(r)$ and
approximating the sum of the integrals by $M^*(T)T$, we solve Eq.
(8) and obtain \beq M^*(T)\propto \frac{1}{\sqrt{T}}.\eeq Thus, we
can conclude that at elevated temperatures when $x\simeq x_{FC}$
the system exhibits three types of regimes: the LFL behavior at
$\alpha\ll 1$, when Eq. (10) is valid; the $M^*(T)\propto
T^{-2/3}$ behavior, when Eqs. (12) and (13) are valid; and the
$1/\sqrt{T}$ behavior of the effective mass at $\alpha\sim 1$.

Since the resistivity, $\rho(T)=\rho_0+\Delta\rho$, with $\rho_0$
being the residual resistivity and $\Delta\rho=A(B)T^2$, is
directly determined by the effective mass because the coefficient
$A(B)\propto (M^*)^2$ \cite{ksch} the above mentioned regimes can
be observed in measurements of the resistivity. The first LFL
regime is related to Eq. (7) and represented by
$\Delta\rho_1=c_1T^2/(B-B_{c0})^{4/3}\propto T^2$; the second NFL
one is determined by Eq. (11) and  characterized by
$\Delta\rho_2=c_2T^2/(T^{2/3})^2\propto T^{2/3}$; and the third
NFL one is given by Eq. (14) and represented by
$\Delta\rho_3=c_3T^2/(\sqrt{T})^2\propto T$. Here $c_1,c_2,c_3$
are constants. It is remarkable that all these regimes were
observed in measurements on the HF metals
\cite{bud,pag1,pag,movsh}. If we consider the ratio
$\Delta\rho_2/\Delta\rho_1\propto ((B-B_{c0})/T)^{4/3}$, we arrive
at the very interesting conclusion that the ratio is a function of
only the variable $(B-B_{c0})/T$ representing the scaling
behavior. This result is in excellent agreement with experimental
facts \cite{pag}.

We now turn to the estimation of the quasiparticles width
$\gamma(T)$. Within the framework of the LFL theory the width is
given by \cite{lanl1} \beq \gamma\sim |\Gamma|^2(M^*)^3T^2,\eeq
where $\Gamma$ is the particle-hole amplitude. In the case of a
strongly correlated system with its large density of states
related to the huge value of the effective mass, the amplitude
$\Gamma$ cannot be approximated by the bare particle interaction
but can be estimated within the ladder approximation which gives
$|\Gamma|\sim 1/(p_FM^*(T))$ \cite{duk}. As a result, we have that
in the LFL regime $\gamma(T)\propto T^2$, in the $T^{-2/3}$ regime
$\gamma(T)\propto T^{4/3}$, and in the $1/\sqrt{T}$ regime
$\gamma(T)\propto T^{3/2}$. We observe that in all the cases the
width is small compared to the quasiparticle characteristic energy
which is of the order of $T$. We can conclude that when the heavy
electron liquid is near the QCP being on the disordered side its
low energy excitations are quasiparticle excitations with the
effective mass $M^*(T,B)$. At this point we note that at $x\to
x_{FC}$, the quasiparticle renormalization factor $Z$ remains
finite and approximately constant, and the divergence of the
effective mass given by Eq. (4) is not related to vanishing $Z$,
see e.g. \cite{khod1}. Thus the notion of the quasiparticles is
preserved and these are the relevant excitations when considering
the thermodynamical properties of the heavy electron liquid.

As we have seen above at $T=0$ when $r=(x-x_{FC})\to 0$, the
effective mass $M^*(r)\to\infty$ and eventually beyond the
critical point $x_{FC}$ the distance $r$ becomes negative making
the effective mass negative as follows from Eq. (4). To escape the
possibility of being in unstable and meaningless states with
negative values of the effective mass, the system is to undergo a
quantum phase transition at the critical point. Because the
kinetic energy near the Fermi surface is proportional to the
inverse effective mass, this phase transition is triggered by the
frustrated kinetic energy and can be recognized as FCQPT
\cite{shag3}. Therefore behind the critical point $x_{FC}$ of this
transition, the quasiparticle distribution function represented by
the step function does not deliver the minimum to the Landau
functional $E[n({\bf p})]$. As a result, at $x<x_{FC}$ the
quasiparticle distribution is determined by the standard equation
to search the minimum of a functional \cite{ks}
\begin{equation} \frac{\delta E[n({\bf p})]}{\delta n({\bf
p},T=0)}=\varepsilon({\bf p})=\mu; \,p_i\leq p\leq p_f.
\end{equation}
Equation (16) determines the quasiparticle distribution function
$n_0({\bf p})$ which delivers the minimum  value to the ground
state energy $E$. Being determined by Eq. (16), the function
$n_0({\bf p})$ does not coincide with the step function in the
region $(p_f-p_i)$, so that $0<n_0({\bf p})<1$, while outside the
region it coincides with the step function. It follows from Eq.
(16) that the single particle spectrum or the band is completely
flat over the region. Such a state was called the state with
fermion condensate (FC) because quasiparticles located in the
region $(p_f-p_i)$ of momentum space are pinned to the chemical
potential $\mu$ \cite{ksk,ks,vol}. We note that the behavior
obtained for the band and quasiparticle distribution functions as
observed within exactly solvable models \cite{irk,lid}. We can
conclude that the relevant order parameter $\kappa({\bf
p})=\sqrt{n_0({\bf p})(1-n_0({\bf p}))}$ is the order parameter of
the superconducting state with the infinitely small value of the
superconducting gap \cite{ksk}. Thus this state cannot exist at
any finite temperatures and driven by the parameter $x$: at
$x>x_{FC}$ the system is on the disordered side of FCQPT; at
$x=x_{FC}$, Eq. (16) possesses the non-trivial solutions $n_0({\bf
p})$ with $p_i=p_F=p_f$; at $x<x_{FC}$, the system is on the
ordered side.

At $T>0$, the quasiparticle distribution is given by Eq. (2) which
can be recast as
\begin{equation} \varepsilon ({\bf p},T)-\mu (T)=T\ln \frac{1-n({\bf
p},T)}{n({\bf p},T)}.  \end{equation} As $T\to 0$, $n({\bf
p},T)\simeq n_0({\bf p})$, the logarithm on the right hand side of
Eq. (17) is finite when $p$ belongs to the region $(p_f-p_i)$,
therefore $T\ln(...)\to 0$, and we again arrive at Eq. (16). It
follows from Eq. (17) that for $r<0$ and $T\to 0$, the effective
mass diverges as \cite{ksk} \beq M^*(T)\sim
p_F\frac{p_f-p_i}{4T},\eeq while the width of the quasiparticles
being given by Eq. (15) is $\gamma\sim T$. We see that  at lower
temperatures and beyond FCQPT, the heavy electron liquid behaves
as if it were placed at QCP, in fact it is placed at the quantum
critical line and Eq. (18) is valid at any $x$ provided that
$x<x_{FC}$. One could suggest that this quantum critical behavior
originates from the paramagnet to superconducting quantum phase
transition and the corresponding quantum fluctuation region
contributes to the thermodynamical properties. Fortunately, this
is not the case since the width in the temperature of the
fluctuating regime is very narrow and practically irrelevant from
experimental point of view \cite{lanl1}. Thus, the quasiparticle
scenario is applicable down to the smallest temperatures. For
example, the LFL and NFL behavior and the scaling properties at
microkelvin temperatures (down to 170 $\mu$K) and ultra small
magnetic fields ($0.02\sim6.21$ mT) recently investigated
experimentally on the HF metal CeRu$_2$Si$_2$ \cite{takah} have
been explained within the FCQPT scenario \cite{shag4,ckhz}.

At this point, we consider how the behavior of the effective mass
given by Eqs. (14) and (18) correspond to experimental
observations. It was recently observed that the thermal expansion
coefficient $\alpha(T)/T$ measured on CeNi$_2$Ge$_2$ shows a
$1/\sqrt{T}$ divergence over two orders of magnitude in the
temperature range from 6 K down to at least 50 mK, while
measurements on YbRh$_2$(Si$_{0.95}$Ge$_{0.05}$)$_2$ demonstrate
that $\alpha/T \propto 1/T$ \cite{geg1}, contrary to  the LFL
theory which yields $\alpha(T)/T\propto M^*\simeq const$. Since
the effective mass depends on $T$, we obtain that the $1/\sqrt{T}$
behavior, Eq. (14), is in excellent agreement with the result for
the former system \cite{alp}, and the $1/T$ behavior, Eq. (18),
predicted in \cite{zver} corresponds to the latter HF metal.
Obviously, a theory based on the critical fluctuations related to
CQPT cannot explain the observed behavior of these divergences
taking place at least over one order of magnitude in temperature
change. One could expect that a temperature scale $\tau$ exists
where the quantum fluctuations dominate the behavior of the heavy
electron liquid controlling the singular contribution to the free
energy at $T<\tau$. There are experimental facts collected on the
magnetization and magnetic susceptibility of the HF metal
CeRu$_2$Si$_2$ with $B_{c0}=0$ \cite{takah}. These demonstrate
that the application of tiny magnetic fields of $0.2$ mT at
microkelvin temperatures of $200$ $\mu$K restores the LFL behavior
related to the presence of quasiparticles and therefore destroying
the quantum fluctuations. At temperatures of as high as $3$ mK
these fluctuations survive to destroy the LFL behavior. However,
these are destroyed by magnetic fields of $0.94$ mT thereby
restoring the LFL behavior. Then, the measured susceptibility and
magnetization show the scaling behavior down to the lowest
temperatures. On the other hand, if the scale exists the scaling
behavior related with the quasiparticle contribution to the free
energy would fail at $T<\tau$. In fact, the width of the regime of
the fluctuations is narrow and can hardly take place even over one
order of magnitude in temperature change \cite{lanl2}. In contrast
the quasiparticle scenario and the observed behavior make a
perfect match. Thus, we conclude that the Landau paradigm based on
the notions of the quasiparticles and order parameter is
applicable when considering the heavy electron liquid.

At $T=0$, the application of a magnetic field $B$ splits the FC
state into the Landau levels and suppresses the superconducting
order parameter $\kappa({\bf p})$ destroying the FC state.
Therefore the LFL behavior is expected to be restored \cite{shag}.
The Landau levels at the Fermi surface can be approximated by a
single block whose thickness in momentum space  is $\delta p$.
Approximating the dispersion of quasiparticles within this block
by $\varepsilon(p)\sim (p-p_F+\delta p)(p-p_F)/M$, we obtain that
the effective mass $M^*(B)\sim M/(\delta p/p_F)$. The energy loss
$\Delta E_{FC}$ due to rearrangement of the FC state related to
this block  can be estimated using the Landau formula \cite{lanl1}
\begin{equation}
\Delta E_{ FC}=\int(\varepsilon({\bf p})-\mu)\delta n({\bf
p})\frac{d{\bf p}^3}{(2\pi)^3}.
\end{equation}
The region occupied by the variation $\delta n({\bf p})$ has the
length $\delta p$, while $(\varepsilon({\bf p}) -\mu)\sim
(p-p_F)p_F/M^*(B)$. As a result, we have $\Delta E_{ FC}\sim\delta
p^2/M^*(B)$. On the other hand, there is a gain $\Delta E_B\sim
(B^2\mu_{B})^2M^*(B)p_F$ due to the application of the magnetic
field and coming from the Zeeman splitting. Equating $\Delta E_B$
to $\Delta E_{FC}$ and taking into account that in this case
$M^*(B)\propto 1/\delta p$, we arrive at the following relation
\begin{equation} \frac{\delta
p^2}{M^*(B)}\propto \frac{1}{(M^*(B))^3}\propto B^2M^*(B).
\end{equation} It follows from Eq.
(20) that the effective mass $M^*(B)$ diverges as
\begin{equation} M^*(B)\propto \frac{1}{\sqrt{B-B_{c0}}}.
\end{equation}
Here again we have substituted $B$ by $(B-B_{c0})$ as it was done
when deriving Eq. (7). Equation (21) shows that by applying the
magnetic field $B>B_{c0}$ the system can be driven back into the
LFL with the effective mass $M^*(B)$ depending on the magnetic
field. This means that the coefficients $A(B)\propto (M^*(B))^2$,
the specific heat, $C/T=\gamma_0(B)\propto M^*(B)$, and the
magnetic susceptibility $\chi_0(B)\propto M^*(B)$. It is seen that
the well-known empirical Kadowaki-Woods (KW) ratio \cite{kadw},
$K=A/\gamma_0^2\simeq const$, is obeyed. At this point, we stress
that the value of $K$ may be dependent on the degeneracy number of
quasiparticles. In the simplest case when the heavy electron
liquid is formed by quasiparticles with the spin $1/2$ and the
degeneracy number is $2$, $K$ turns out to be close to the
empirical value \cite{ksch}, called as the KW relation
\cite{kadw}.

It follows from Eq. (21) that the coefficient $A(B)$ diverges as
\beq A(B)\propto \frac{1}{B-B_{c0}}.\eeq We note that in contrast
to the LFL the effective mass strongly depends on the magnetic
field diverging at $B-B_{c0}$. Such a behavior resembles the
behavior when the heavy electron liquid approaches a field tuned
QCP but in the considered case the system exhibits the behavior at
any point $x<x_{FC}$, thus, in fact we are dealing with the
critical line. The divergence of the effective mass at QCP is
described by Eq. (7) and the coefficient $A(B)$ diverges as \beq
A(B)\propto \frac{1}{(B-B_{c0})^{4/3}}.\eeq While at the critical
line, the divergence of the effective mass and the coefficient
$A(B)$ are given by Eqs. (21) and (22) respectively. The observed
two types of divergences and the constancy of the KW ratio are in
excellent agreement with recent experimental facts collected on
the HF metals YbRh$_2$Si$_2$ \cite{geg},
YbRh$_2$(Si$_{0.95}$Ge$_{0.05}$)$_2$ \cite{cust}, YbAgGe
\cite{bud} and CeCoIn$_{5}$ \cite{pag1,pag,movsh}. Thus we are led
to the conclusion that it is possible to describe two different
types of behavior related to the different exponents based on the
single quantum phase transition. This is the distinctive feature
of FCQPT which is in good agreement with recent measurements on
the HF metals. Because the effective mass is defined by the
magnetic field, Eq. (21), it is possible to control by the
application of magnetic field the main thermodynamical properties
of the heavy electron liquid at $x<x_{FC}$.

To analyze the behavior of the system at rising temperatures we
use Eq. (5).  Elevated temperatures change the LFL behavior
induced by the magnetic field $B$ into the NFL behavior. Using the
same arguments which led to Eq. (11), we see that the heavy
electron liquid demonstrates the $M^*(T) \propto T^{-2/3}$ regime
provided that high magnetic fields are applied. Finally at
$T^*(B)<T_k$, the behavior of the effective mass is given by Eq.
(18). To estimate $T^*(B)$ characterizing the crossover region, we
have to equate the effective mass $M^*(T)$ defined by Eq. (18) to
$M^*(B)$ given by Eq. (21), $M^*(T)\sim M^*(B)$. As a result, we
obtain
\begin{equation} T^*(B)\propto \sqrt{B-B_{c0}}.
\end{equation}

In summary, we have shown that the Landau paradigm is still
applicable when considering the low temperature properties of the
heavy electron liquid, whose understanding has been problematic
largely because of the absence of theoretical guidance. In
contrast with the conventional Landau quasiparticles, the
effective mass of the considered quasiparticles in this letter
strongly depends on the temperature and the applied magnetic
field, while the order parameter is destroyed at any finite
temperature. These quasiparticles and the order parameter are well
defined and capable of describing both the LFL and the NFL
behaviors of the HF metals and their universal thermodynamic
properties down to the lowest temperatures. We have shown that
even such a subtle behavior as the recently observed anomalous
$T^{2/3}$ dependence of the resistivity and the corresponding
scaling behavior can be understood. Additionally, we have
demonstrated that this unusual behavior of both the order
parameter and the quasiparticles is determined by FCQPT which
allows the existence of the quasiparticles down to the lowest
temperatures. In that case we obtain a unique possibility to
control the essence of the HF
metals by magnetic fields in a wide range of temperatures.\\

\bigskip \noindent{\bf Acknowledgements}\\

The visit of VRS to Clark Atlanta University has been supported by
NSF through a grant to CTSPS. MYaA is grateful to the S.A.
Shonbrunn Research Fund for support of his research. AZM is
supported by US DOE, Division of Chemical Sciences, Office of
Basic Energy Sciences, Office of Energy Research. This work was
supported in part by the the National Science Foundation under Grant 
No. PHY99-07949 and by the Russian Foundation for Basic Research.

\end{document}